# Experimental investigation of delayed self-interference for single photons

Nicklas Ohlsson, Mattias Nilsson and Stefan Kröll

Department of Physics, Lund Institute of Technology (LTH), Box 118, S-221 00 Lund, Sweden

**Abstract**

A photon echo experiment has been performed using accumulated highly attenuated laser pulses. We show experimentally that the photon echo process can be performed with, on the average, less than one photon in each pair of excitation pulses. The results support an interpretation where this non-linear process can be performed with, on the average, less than one photon shared between two of the optical fields involved in the degenerate four-wave mixing process. Further, we argue that the experiment can be interpreted as a form of delayed self-interference for photon wave packets that do not overlap in both space and time.

**Introduction**

In 1993, Kessel and Moiseev [1] proposed an experiment for observation of delayed self-interference for single photons. Their idea was to split the wave packet of a single photon using a 50 % beam-splitter and to let the two paths from the beam splitter overlap in space, such that the difference in optical path-length between the two paths exceeds the length of the single-photon wave packet, see Fig. 1. If a normal detector, e.g. a photographic plate, is placed where the two paths overlap, no interference pattern will be visible, in accordance with the notion that the wave packets must overlap in both space and time for interference to take place. However, Kessel and Moiseev proposed that if instead of the normal detector, a photon echo material with a combination of long phase memory and inhomogeneously broadened absorption profile is placed at the overlap region, interference between the wave packets travelling the two possible paths would take place.

Materials appropriate for this experiment include e.g. inorganic crystals doped with rare-earth ions, organic materials doped with dye molecules and atomic gases experiencing Doppler broadening. In the proposed experiment, the delayed interference will be manifested in two ways. The angle between the two paths will give rise to a spatially varying modulation of the population of absorbing atoms in the photon echo material and the time delay between the wave packets travelling along the two paths will result in a periodic modulation of population in the frequency domain. After illumination with a strong optical pulse, which in the present paper will be denoted the read-out pulse, the frequency dependent modulation of population will give rise to a coherent emission of a pulse of light, a photon echo, after a time exactly matching the time difference between the wave packets travelling along the two possible paths. The direction of this emission will be given by the phase-matching conditions for the photon echo process.

If only a single photon is made to interfere in the sample before the read-out pulse is sent in, it has been argued that interference will still take place, and that this interference could, in principle, be detected [1,2]. The photon echo emitted by the material as a response to the read-out pulse in this case has been predicted to have new and interesting quantum optical properties. It is clear that the frequency dependent modulation of the population in the medium after interaction with a single photon will be very weak, and therefore the signal indicating the existence of this modulation in the medium would be very difficult to detect. In fact, it has been shown that the expectation value of the intensity of the signal in this case is zero. However, the expectation value of the amplitude of the optical field is non-zero and could therefore, in principle, be detected using phase-sensitive detection. In order to increase the strength of the signal from the medium, it has been suggested [1,3] that one could use so called accumulated photon echoes [4]. In this case (Fig. 2), several single photon events would coherently add to each other to form a strong frequency dependent modulation of the population in the medium. The echo intensity is proportional to the square of the population modulation and this non-linear nature of the photon echo process makes the coherent accumulation of the single photon events more efficient than accumulating the signals after read-out. Coherent accumulation of pulse pairs in order to build up a strong echo signal in rare-earth-ion doped crystals has been studied before, but without attempting to use excitation pulses with energies at the single photon level [5]. It should be noted that the accumulation of many single-photon events before read-out, resulting in non-zero photon echo intensity, will wash out the peculiar non-classical properties of the signal mentioned above.

In the present paper, we present an experimental investigation of a modified version of the delayed single-photon self-interference experiment described above. Due to experimental advantages that will be discussed later in this paper we have, instead of splitting the wave packets with a beam splitter and having two spatially separated paths to the sample as described above, used a collinear geometry for the excitation pulses [6]. This means that only the temporal, and not the spatial, part of the delayed single-photon interference effect will be manifested in our experiment. The experiments were performed using a sample in the form of an inorganic crystal doped with rare-earth ions, namely $Y_2SiO_5$ doped with praseodymium, immersed in liquid helium. Attenuated pulses from a dye laser with energies down to what corresponds to, on the average, 0.54 photons per pulse pair, gave detectable echo signals.

Because of the non-linear nature of the photon echo process and the fact that the interactions take place at two distinct moments in time, one might think that there must be at least one photon present in each of the excitation pulses that build up the population modulation in the material. We have therefore compared our data with theoretical models predicting the behaviour of the signal as a function of the average number of photons per pulse pair both for the case when the delayed single-photon self-interference effect exists and the case when at least two photons are needed in each pulse pair. Our conclusion is that our data strongly suggests that the process described above can be performed by a single photon shared between the two pulses.

The experiment presented in the present paper is an interference experiment where two wave packets in the time domain interfere to produce an interference pattern in the frequency domain. Self-interference for single photons in the spatial domain, as in Young's double slit experiment, has been investigated experimentally, clearly showing the possibility for single photons to produce a spatial interference pattern [7]. Under certain conditions, the absorption of two consecutive optical pulses can be regarded as a time analogue of the double slit experiment. Several photon echo experiments have pointed at this type of analogies between coherent effects in space and time [8,9]. However, the present experiment is, to the best of our knowledge, the first such experiment where the quantisation of the electromagnetic field has to be considered. The use of inhomogeneously broadened absorbers for exploring non-stationary quantum optics in ways different from the experiment presented in this paper has also been given attention lately [10,11]. In the present experiment, optical fields resonant with only one optical transition are used, meaning that only a two level system is utilised. Experiments using optical pulses resonant with different optical transitions in three level systems have also been proposed [12].

The paper is organised in the following way. First a description of the details of the experiment is given, including motivations for the choice of material and equipment used in the experiment. Then the experimental results are presented and compared to theoretical predictions. Finally, a discussion about the conclusions that can be drawn from this experiment is given. The paper ends with a summary.

## Experiment

In the original proposal for the delayed single-photon self-interference experiment, by Kessel and Moiseev, a single photon wave packet was to be split by a beam-splitter [1]. The two possible paths for the wave packets were then proposed to have different optical path lengths before they were to be overlapped in space at the sample. Our experiment differs from that proposal in two aspects; the use of highly attenuated laser pulses instead of true single-photon states and the choice of a collinear geometry for the wave packet paths.

Experiments with true single-photon states are non-trivial to realise, especially considering the demands on repetition rate, wavelength and pulse duration that this experiment would put on a single photon source. Lately, substantial progress has been made in developing and using single photon light sources e.g. Ref. [13], but as in many other experiments exploring properties of single photons [14,15], attenuated laser pulses have been used in the present experiment. By taking the photon statistics in such pulses into account, it is still possible to deduce effects due to single photons.

The original proposal used two spatially separated photon paths to the sample to create the time delay between the two wave packets of the photon and an angle between these paths to create a spatially varying population modulation in the material. Experimentally this is non-trivial to realise. In order for all the single photon events to accumulate coherently in the photon echo material, the variation in optical path length difference between the two paths must be stabilised to significantly below one wavelength of the light used and this stability must be maintained during the whole accumulation sequence. In our case, the longest accumulation time used was almost one hour and involved the accumulation of more than $10^9$ pulse pairs. To maintain the required stability during such a long accumulation sequence complicates the experiment. To overcome the complication regarding stabilisation of the path lengths, we have instead used a collinear geometry for the light in our experiment. Using acousto-optic modulators, we create two pulses from the continuous laser source. These two pulses correspond to the two possible photon paths in Ref. [1]. The time difference between the two pulses mimics the difference in optical path length. Using this scheme, the requirement of stability of the interferometer in the original proposal is transferred to a requirement that there is a constant phase difference between the two pulses for all the accumulated pulse pairs. Practically this means that the maximal time between two pulses that can be chosen is of the order of the coherence time of the light source used for the experiment.

Detailed calculations of the dependence of the signal on different parameters for the photon echo material have been performed earlier [3]. The calculations show that if the effect of all pulse pairs add coherently, the number of photons, *S*, in the echo should scale as:

$$S \propto N^2 \cdot M^2 \cdot T, \qquad (1)$$

where *N* is the number of photons in each excitation pulse pair, *M* is the number of pulse pairs that are accumulated before read-out and *T* is the duration of the excitation pulses. Besides the above, which is only a result of the non-linear nature of the photon echo process, the calculations showed that a material with a large value of the oscillator strength for the optical transition should be chosen. However, the material must still have a phase memory considerably longer than the time separation between the excitation pulses in each pulse pair for the accumulation to work efficiently. It is also desirable to have a material with a high efficiency for storing long-lived spectral features. The interference between the pulses in each pulse pair creates a frequency dependent modulation of the population in both the ground- and excited states. If the population in the excited state decays to a state different from the ground state, the modulation of the ground state population will still be preserved. The percentage of ions decaying to a state different from the ground state is a measure of the material's efficiency for storing long lived spectral features. The lifetime for the storage mechanism, which for the rare-earth-ion doped crystals could be e.g. the relaxation time between the hyperfine levels of the ground state, puts an upper limit on the accumulation time that can be used in the experiment. It is therefore also desirable to have a material where this mechanism is as long lived as possible. Further, the photon echo signal is optimised when the length and concentration of absorbers of the sample is chosen so that the optical density for the sample is of the order of unity. In the present experiment, for the echo signal to be maximised, the read-out pulse should be chosen to have a pulse area of $\pi/2$ for the absorbers for which it is resonant. Under other conditions, this is not necessarily the case [16].

Several different materials have been considered for the experimental realisation. Besides optimising the parameters mentioned above one has to take several other practical issues related to the choice of material into account. A material with a high oscillator strength and thereby with a short phase memory would mean that short pulses and small time separations between the pulses in the pulse pairs must be used for the coherent accumulation to work efficiently. However, if pulses with durations in the picosecond regime are to be used, it is (at present) necessary to use mode-locked lasers instead of externally modulated continuous-wave lasers. The use of such lasers usually means that it is more difficult to tailor pulse durations and repetition rates in order to optimise the signal. The use of small time separations will also make the detection of the signal more difficult, since the separation between the strong read-out pulse and the weak signal will be the same as the separation between the pulses in the accumulated pulse pairs. If an exceedingly short time separation is chosen, the use of an electronically controlled gate to prevent the detector from being saturated by the read-out pulse but still be fully sensitive to the echo signal, will be difficult.

For the rare-earth ion doped crystals, mainly two different dopants have been considered for the experiment, namely europium and praseodymium. The hour-long relaxation time of the hyperfine levels for many europium-doped crystals, e.g. Eu:Y$_2$SiO$_5$, is advantageous since it enables accumulation times of up to several hours. However, the low values of the oscillator strengths (~$10^{-8}$) for the optical transitions in these materials make them unsuited for this experiment. Crystals doped with praseodymium generally have 1-2 orders of magnitude higher value of the oscillator strength. However, the relaxation times for the hyperfine levels in these crystals are much shorter, e.g. a few hundred seconds for Pr:Y$_2$SiO$_5$ [17]. For some rare-earth ion doped crystals the application of an external magnetic field leads to an increase of the lifetime for spectral features stored in the ground state population [18]. We therefore chose to use a sample of Pr:Y$_2$SiO$_5$ placed in an external magnetic field for our experiment and due to the magnetic field we could observe a significant increase of the lifetime for spectral features stored in the ground state.

The experimental investigations were performed on the $^3H_4$-$^1D_2$ transition of praseodymium ions doped into yttrium silicate (Y$_2$SiO$_5$) in site 1, absorbing at 605.8 nm. In this host, praseodymium substitutes for yttrium. Our sample was a 0.5 mm thick crystal with a doping concentration of 0.05 at. %. The crystal was immersed in liquid helium and kept at a temperature of 4.2 K in a cryostat. In Fig. 3, the experimental set-up for the accumulated photon echo experiment can be seen. In the experiment, an argon ion pumped dye laser (Coherent 699-21) was tuned to 605.8 nm with the aid of a home built wave meter. 40 % of the laser light was sent to a locking set-up to obtain an improved long-term frequency stability of the dye laser [19]. This stabilisation was accomplished using Doppler-free saturation spectroscopy to reveal the hyperfine lines of the molecular iodine and locking the laser to one of the lines. The remaining 60 % of the light was sent to three acousto-optic modulators (Isomet 1205) placed in series. The modulators were used for creating the excitation pulses. Three modulators placed in series yielded an extinction ratio better than 1:$10^7$ for the excitation light when the modulators were switched off. In particular this was important because any leakage of light after the read-out pulse would constitute a background at the detector when the echo arrives. Two of the modulators were arranged so that the frequency of the light could be changed by changing the frequency of the RF signal controlling the modulators, without changing the direction of the transmitted light. After the modulators, ND filters were inserted to attenuate the energy of the light in the excitation pulses to the desired level. The light beam (~2 mm diameter) was then focused using a lens with a 15 cm focal length onto the crystal in the cryostat. In order to have a long relaxation time for the hyperfine levels of the praseodymium ions in the crystal, a magnetic field was applied to the crystal as mentioned above. Magnetic coils were positioned in Helmholtz configuration and fed with 2.5 A of current, producing a magnetic field of the order of 0.01 T at the sample. After the cryostat, two acousto-optic modulators in series were used as a gate to block the excitation pulse, thereby preventing the detector from being saturated by the read-out pulse. These modulators were opened immediately after the read-out pulse, allowing only the echo signal to reach the detector. The signal was detected with a photo-multiplier tube (Hamamatsu R 943-02) and monitored on an oscilloscope (Tektronix TDS 540).

With the three acousto-optic modulators fully opened, the power at the cryostat was 28 mW. In the experiment, excitation pulses with durations of 44 ns, separated by $\tau=175$ ns, were used. In order for the read-out pulse to cover the same spectral region of the inhomogeneous absorption line of the crystal as the pulse pairs in the accumulation sequence, the duration of the read-out pulse was the same as that of the accumulated pulses. The different pulse pairs being accumulated were separated by $\sigma=470$ ns, corresponding to a repetition rate of 2.1 MHz.

The energy of the pulses was estimated by first measuring the average power of the accumulation pulses without attenuation, using a Coherent Fieldmaster power meter (model LM-2). By measuring the time separation between the pulse pairs accurately, the energy in each pulse pair could be calculated. Using the same power meter and laser, the attenuation of the ND filters used in the experiments was then measured.

The absorption of two non-saturating pulses of light within the phase memory time of the material will lead to a frequency dependent periodic modulation of the population of absorbers in the material. The periodicity of this spectral modulation is determined by the time difference between the two pulses, such that the frequency difference between two maxima in the modulation is exactly the inverse of the time between the two applied pulses. If several pairs of pulses, with the same time difference between the pulses within the pairs, are sent into the material, each pair will create a frequency dependent modulation with the same periodicity. For all of these modulations to accumulate constructively, i.e. for the maxima of the modulations to overlap in frequency, two things have to be fulfilled. First, the phase difference between the light in the two pulses in each pair has to be the same for all pulse pairs. This means that the laser used in the experiment must have a coherence time that is longer than the time separation τ between the pulses within a pulse pair. Second, the variation in the frequency of the light for different pulse pairs should be smaller than the frequency difference between the maxima of the population modulations. This means that the frequency of the laser, during the whole accumulation sequence, has to stay within a frequency interval which is considerably smaller than that corresponding to the inverse of the time difference between the pulses in the pulse pairs. The dye laser that was used in our experiments has a line-width of approximately 1 MHz, meaning a coherence time long enough for the coherent accumulation of pulse pairs with the pulse separation of 175 ns that was used in the experiment. However, we could clearly observe that the accumulation efficiency decreased when the pulse separation was increased. The long time stability of the laser was improved using the iodine locking mentioned above. When the locking was used, the frequency of the laser could be kept within a few MHz for several hours [19].

To obtain a pulse area of the read-out pulse as close as possible to π/2, all ND filters were removed from the optical path before read-out. The short duration of the pulse in combination with the limited output power from the laser still meant that the pulse did not quite have the desired pulse area, leading to a reduction of the photon echo signal. On the other hand, this allowed several read-outs without destroying the frequency dependent modulation of population in the material, thus making averaging of several echo signals possible.

To measure the relaxation time of the hyperfine levels in the applied magnetic field, a spectral hole burning experiment [20] was performed. The same optical components were used in the set-up as those described above, but without focusing the light onto the crystal and with only a photodiode on the detection side. A pulse with a duration of 20 ms was applied to the crystal. This pulse was used to burn a spectral hole in the inhomogeneous absorption profile of the crystal. By repeatedly scanning the frequency of the laser over the spectral region where the hole was burnt and monitoring the area of the hole, the relaxation time of the hyperfine levels can be estimated. The laser power when monitoring the hole was approximately 100 µW and the duration of these pulses was 100 µs. This was low enough not to destroy the spectral hole even after a large number of monitoring pulses. The spectral hole was monitored every 20 s and the resulting relaxation curve can be seen in Fig. 4. For comparison, the relaxation curve in zero magnetic field was obtained and is shown in the same figure. In this case the spectral hole was monitored every 5 s. It is known that the hyperfine population relaxation in Pr:$Y_2SiO_5$ cannot be described by a single exponential decay [17]. In zero magnetic field, we could observe a fast component with a relaxation time of ~100 s and a more long-lived component with a decay time of several hundred seconds. With an applied magnetic field of ~0.01 T we observed a component with a decay time of several thousand seconds.

The longer lifetime of the spectral features that was obtained when the magnetic field was applied meant that longer accumulation sequences could be employed before read-out. However, it also made consecutive experiments more difficult since the photon echo signal from previous experiments would not disappear by itself through relaxation. Therefore, after each experimental sequence of accumulation and read-out, the crystal was exposed to a 5 s long pulse with approximately 50 µW of power. During this pulse, the frequency of the light was repeatedly scanned over approximately 80 MHz every 500 µs. The purpose of this procedure was to repeatedly excite all the ions in the crystal that had been affected by the accumulation sequence. After each excitation, the ions would then randomly end up in one of the hyperfine levels of the ground state, thereby erasing the modulation created by the accumulated pulse pairs. This allowed us to perform several consecutive experiments under identical conditions, without having to wait for the population to relax between the hyperfine levels.

## Results and comparison with theory

In Fig. 5b the resulting photon echo signal after accumulation of pulse pairs, each containing on the average 0.54 photons, can be seen. A total of $7.1*10^9$ pulse pairs were accumulated before read-out and the presented data is the average of 5 consecutive read-outs after a single accumulation. To make sure that the observed signal was really the result of the accumulated pulse pairs, a read-out was performed before the accumulation sequence. This read-out, as can be seen in Fig. 5a, did not result in any observable signal. The pulse that can be seen to the left in both the traces is part of the strong read-out pulse leaking through the two acousto-optic modulators that were used as a gate. This gate suppressed the light enough to prevent the read-out pulse from saturating the detector.

Since the pulses used in our experiment were attenuated pulses from a laser, and not true single-photon states, there will always be a certain probability for the pulse pairs to contain more than one photon. The laser field from a laser working far above threshold can be described as a coherent state of the quantised electromagnetic field, containing a photon number with Poissonian distribution [21]. This means that the probability, $P_n$, of finding $n$ photons in a pulse, where the average number of photons is $N$, is given by:

$$P_n = \frac{N^n \cdot e^{-N}}{n!}. \qquad (2)$$

The scaling of the signal with the number of pulse pairs and the number of photons in each pulse pair that was given in Eq. (1) is valid only if the delayed single-photon self-interference effect exists. If the effect does not exist, the scaling has to be modified to take into account that pulse pairs containing only one photon will not contribute to the echo signal. The derivation of the dependence of the signal on the average number of photons per pulse pair is similar to that given in Ref. [2].

According to Eq. (1), the echo signal, due to the accumulation of pulse pairs, should only depend on the total amount of energy used in these pulse pairs, proportional to the product of the number of photons per pair and the number of such pairs. This means that if accumulation of a certain number of pulse pairs each containing a certain amount of energy gives a detectable echo signal, this signal should be the same if instead twice the number of pulse pairs, attenuated by a factor of two are accumulated. However, if the delayed single-photon self-interference effect does not exist, i.e. if at least two photons are needed in each pulse pair to produce an echo signal, the signal would start decreasing when the pulse pairs are attenuated to the level where there is a non-negligible probability for the pulse pairs to contain less than two photons. This means that by observing the behaviour of the echo signal using a fixed amount of energy in the accumulation sequence, but distributing this energy over an increasing number of pulse pairs, it will be possible to conclude whether the delayed single-photon self-interference effects exists or not.

Taking the photon statistics of the laser pulses into account and using Eq. (1), the number of photons in the echo signal, $S_{total}$, should scale as:

$$S_{total} \propto \left( \sum_{n=1}^{\infty} n \cdot P_n \right)^2 \cdot M^2 = N^2 \cdot M^2, \qquad (3)$$

as expected if all pulse pairs are assumed to contribute to the echo signal. However, if the delayed single-photon self-interference effect is assumed not to exist and only pulse pairs containing at least two photons contribute to the signal, the number of photons in the echo signal, $S_{n \neq 1}$, should instead scale as:

$$S_{n \neq 1} \propto \left( \sum_{n=2}^{\infty} n \cdot P_n \right)^2 \cdot M^2 = N^2 \cdot M^2 \cdot \left(1 - e^{-N}\right)^2. \qquad (4)$$

Explicitly requiring one photon in each pulse instead of two photons per pulse pair is an even stricter constraint and would lead to the prediction of an even smaller signal. A series of measurements was performed where the total amount of energy in the accumulated pulse pairs was held fixed, but the number of such pairs (and therefore also the amount of energy in each pair) was varied. First an accumulation of $7.1*10^9$ pulse pairs each containing an energy corresponding to 0.54 photons were accumulated and the signal was monitored. Then the ND filters were changed to give an energy of the pulse pairs corresponding to 1.65 photons per pulse pair and the accumulation time was changed to give fewer accumulated pulse pairs but the same total amount of energy for the accumulation sequence. In the same way, pulse pairs with increasing energy were accumulated and the resulting echo signal was monitored.

Figure 6 shows two series of measurements where the areas of the resulting echo signals have been plotted as a function of the average number of photons in the excitation pulse pairs. In the same figure, the normalised curves corresponding to the behaviour expected from theory for the case where the delayed single-photon self-interference effect exists, Eq. (3) and the case if it does not exist, Eq. (4), have been drawn (solid and dashed traces respectively). The experimental data has been normalised to the average value of the signals resulting from pulse pairs containing an average of 12.5 photons per pulse pair (the experimental points furthest to the right in the figure). In our experiment, lower photon number per pulse pair also meant that a longer time was used for the accumulation sequence and the effects of relaxation during the accumulation sequence, which could last up to one hour, have to be taken into account. As mentioned, the spectral modulations of the population that is a result of the accumulation of pulse pairs, is stored in the population distribution of the hyperfine levels of the ground state of the praseodymium ions. The relaxation of population (c.f. Fig. 4) will lead to a reduced echo signal, especially for longer accumulation sequences, i.e. the measurements with lower average number of photons per pulse pair. In addition to relaxation there are other effects, such as laser frequency shifts and vibrations, that could give smaller signals for decreasing photon numbers (longer accumulation times). However, it should be noted that, even without any compensation for these effect, all the points for lower number of photons per pulse pair are above the dashed curve, which indicates the maximum signal strength if the delayed single-photon self-interference effect does not exist.

The life-time time data shown in Fig. 4 can be used to compensate for relaxation, since the photon echo signal should be proportional to the square of the remaining population modulation at the time of the read-out. A bi-exponential decay curve gave a reasonable fit to the data in Fig. 4. If the fitted curve is integrated for times corresponding to the accumulation times this can be used to compensate for the fact that population modulations created at different times during the accumulation will have relaxed by different amounts at the time of read-out.

In Fig. 7 the data points in Fig. 6 has been compensated for the hyperfine population relaxation and the compensation more than accounts for the decrease in signal for longer accumulation times (lower average number of photons). It is not clear why the calculations described above overcompensates the data for relaxation. One reason could be that the spectral hole studied in the relaxation measurements and the more complex modulation of population created in the echo experiments may not decay at the same rate.

## Conclusions

Although the results obtained in the present experiment follow the theoretical predictions, we believe that there are certain aspects that may be given particular consideration.

The photon echo process is considered to be a degenerate four-wave mixing effect, which is generally phrased as a multi-photon process. In the present case however, less than one photon (on the average) is shared between two of the fields in the process. Even if a pulse pair contains only a single photon, a field amplitude will still be associated with each of the pulses. This indicates that the phrasing multi-field effect is more appropriate for processes of this type. Parametric non-linear optical processes are generally also characterised by conservation of momentum, which follows from (or is an analogue formulation of) the phase matching process. With one photon shared between two of the fields, it is not immediately obvious how momentum is conserved in the present non-linear process. In our opinion, the present experiment supports a point of view that sometimes it is more appropriate to look at the phase matching condition as a pure wave propagation effect.

In the experiment interference between wave packet pairs sharing (on the average) less than one photon is detected in a sample containing macroscopic numbers of absorbing atoms (ions), even though the wave packets never overlap in time and space. The absorption of a photon is seemingly separated into two different moments in time and the choice of time between the two absorption events is only limited by the phase memory of the absorbers. For our choice of material, with a phase memory of about 200 µs, this can detect interference between pulses that are separated by up to 50 km. However, the long phase memory of the absorbers used to detect the interference implies a high spectral resolution, corresponding to a low temporal resolution. Thus, an alternative view is to regard the two correlated pulses in a pulse pair as parts of a single pulse, with a spectral content that is periodic in the frequency domain (with a period that is equal to the inverse of the time separation between the two parts of the pulse). The individual absorbers will act as narrow-band filters and the ensemble will store the information about this spectrum (or interference pattern) in the form of a population modulation. The information is retrieved in time domain by the read-out pulse, which stimulates the emission of a photon echo.

In summary, we have reported on an experiment where the accumulation of highly attenuated laser pulses resulted in a photon echo. The photon echo was created with, on the average, less than one photon in each pair of excitation pulses. We have argued that the non-linear photon echo process can be performed with one photon acting as two of the optical fields involved in the process. Some implications of the results have been discussed, including an interpretation of the experiment as a time-domain analogue of a Young's double slit experiment.

## Acknowledgements

We are grateful to Prof. Sune Svanberg for critical reading of the manuscript. This work was supported by the Swedish Research Council, the Crafoord Foundation and the European Community Information Society Technologies Program under contract IST-2000-30064.

**Figure captions**

**FIG 1.** In the original idea for the delayed single-photon self-interference experiment, a single photon wave packet was proposed to be split by a beam splitter. The two possible paths to the photon echo material were proposed to have a relative time delay larger than the duration of the photon wave packet. A strong read-out pulse would result in the emission of an echo signal, after a time corresponding to the time difference between the two possible paths from the beam splitter to the material.

**FIG 2.** In an accumulated photon echo experiment, several pairs of weak optical pulses, separated by a time interval, $\sigma$, build up a frequency dependent modulation of the population in the medium. A strong read-out pulse will give rise to the emission of a coherent pulse after a time, $\tau$, which is equal to the time between the pulses within the accumulated pulse pairs.

**FIG 3.** Experimental set-up used in our realisation of the delayed single-photon self-interference experiment. A continuos-wave ring-dye laser is pumped by an argon-ion laser. Pulses are created using three acousto-optic modulators (AOMs) in series. The light is focused onto the crystal, which is immersed in liquid helium inside a cryostat. An externally applied magnetic field is used to increase the storage time for spectral features in the crystal. After the cryostat, two AOMs in series are used as a gate to prevent the read-out pulse from saturating the photo-multiplier tube (PMT) used for detecting the echo signal. A number of mirrors, beam-splitters, lenses and apertures are also shown. The components in the dashed rectangle are used for locking the frequency of the dye-laser system to a hyperfine component in iodine. This improved the long-term frequency stability of the laser.

**FIG 4.** Spectral hole decay in $Pr:Y_2SiO_5$. The circles show the relaxation in a magnetic field of around 0.01 T. The reference curve (dots) show the relaxation in zero magnetic field, up to 750 s after the creation of the hole. The curves do not follow single exponentials. Note for instance the long-lived components in both data sets.

**FIG 5.** Accumulation of pulse pairs containing an energy corresponding to 0.54 photons resulted in a detectable echo signal. The strong pulse that can be seen to the left in both a) and b) is a small fraction of the read-out pulse leaking through the closed gate. After this pulse, the gate opens to allow photon echo signal detection. In a) the read-out pulse was sent into the sample without any prior accumulation of pulse pairs. Such measurements were taken as references and no sign of any echo signal can be seen in the recording. In b), an echo signal appears 170 ns after the read-out pulse, due to the almost hour-long accumulation of pulse pairs containing on the average 0.54 photons.

**FIG 6.** The experimentally obtained echo signals for pulse pairs containing different average number of photons are shown (circles). The number of pulse pairs was adjusted to keep the total excitation energy constant. The maximum expected signal under the assumption that the delayed single-photon self-interference effect exists (solid curve) and the corresponding curve assuming that at least two photons are needed in each pulse pair (dashed curve) are also shown. The models have been normalised to the average of the signals obtained for the pulse pairs containing an average of 12.5 photons, where the predictions of the two models do not differ appreciably. Although hyperfine state relaxation during the accumulation has not been taken into account, the data points are still above the signal expected if the delayed single-photon self-interference effect was absent.

**FIG 7.** Experimentally obtained echo signals for pulse pairs containing different average number of photons (circles), where hyperfine state population relaxation during the accumulation has been taken into account. At an average photon number of 0.54 the accumulation sequence lasted 3300 s. The solid curve corresponds to the model where delayed single-photon self-interference contribute to the signal and the dashed curve corresponds to the model where this contribution is assumed to be non-existing.

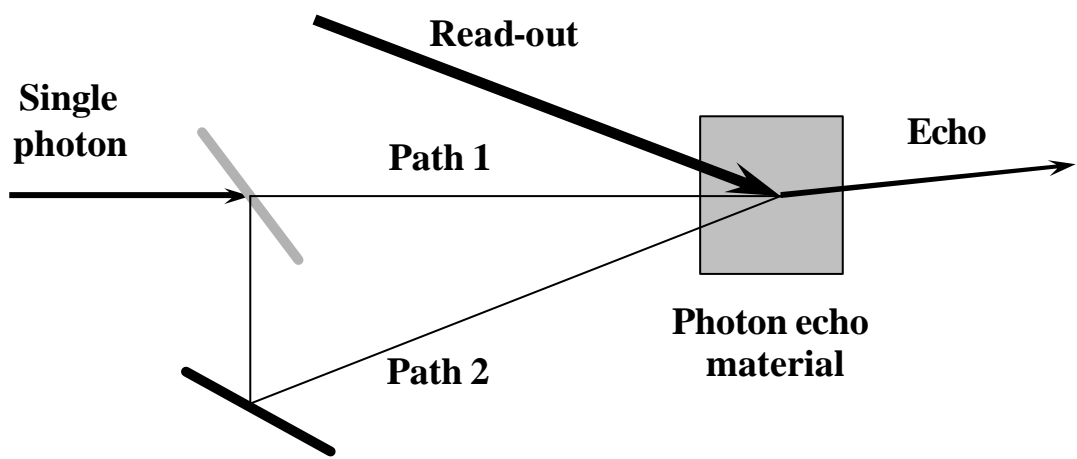

FIG 1

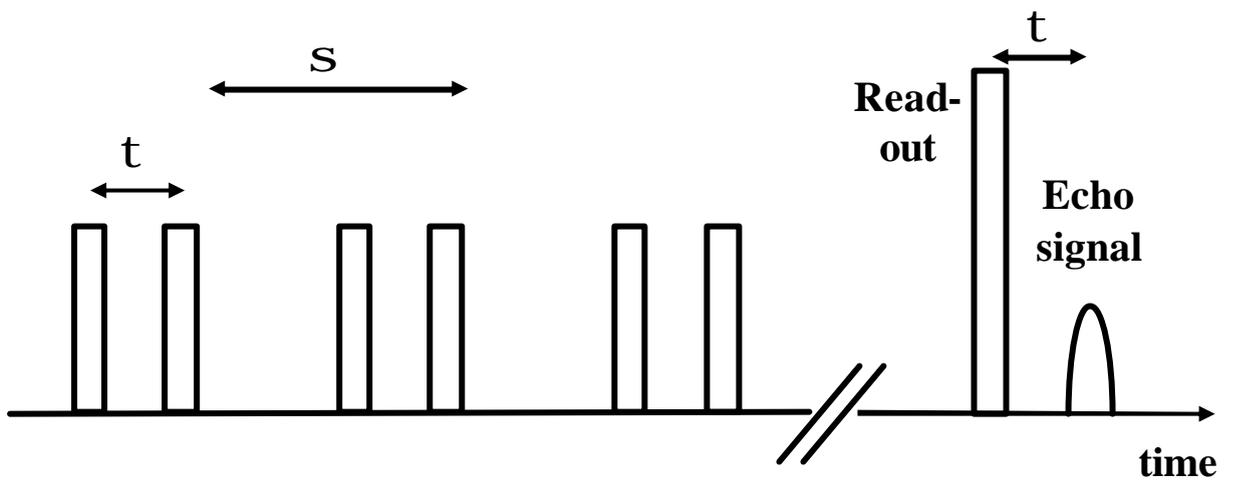

FIG **2**

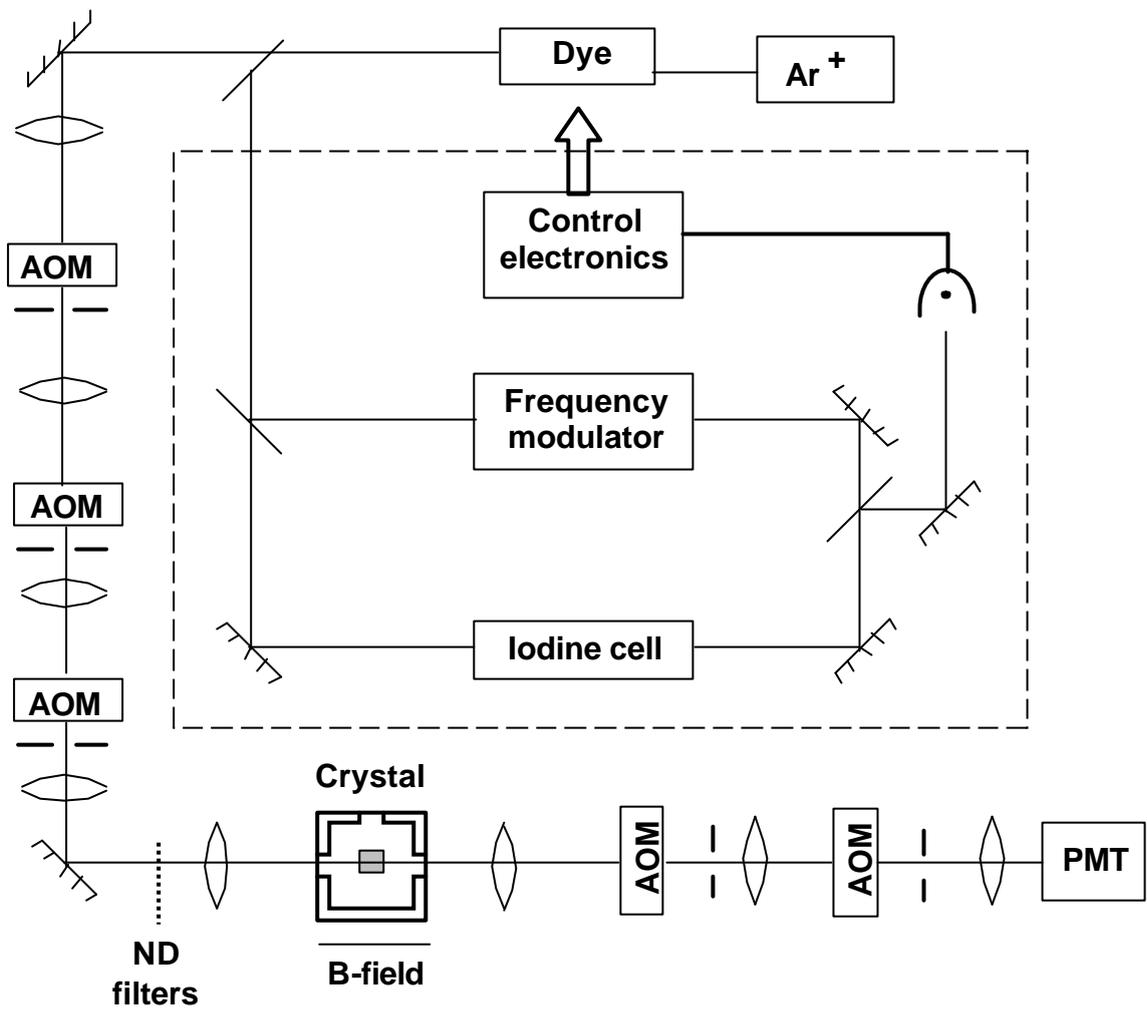

FIG **3**

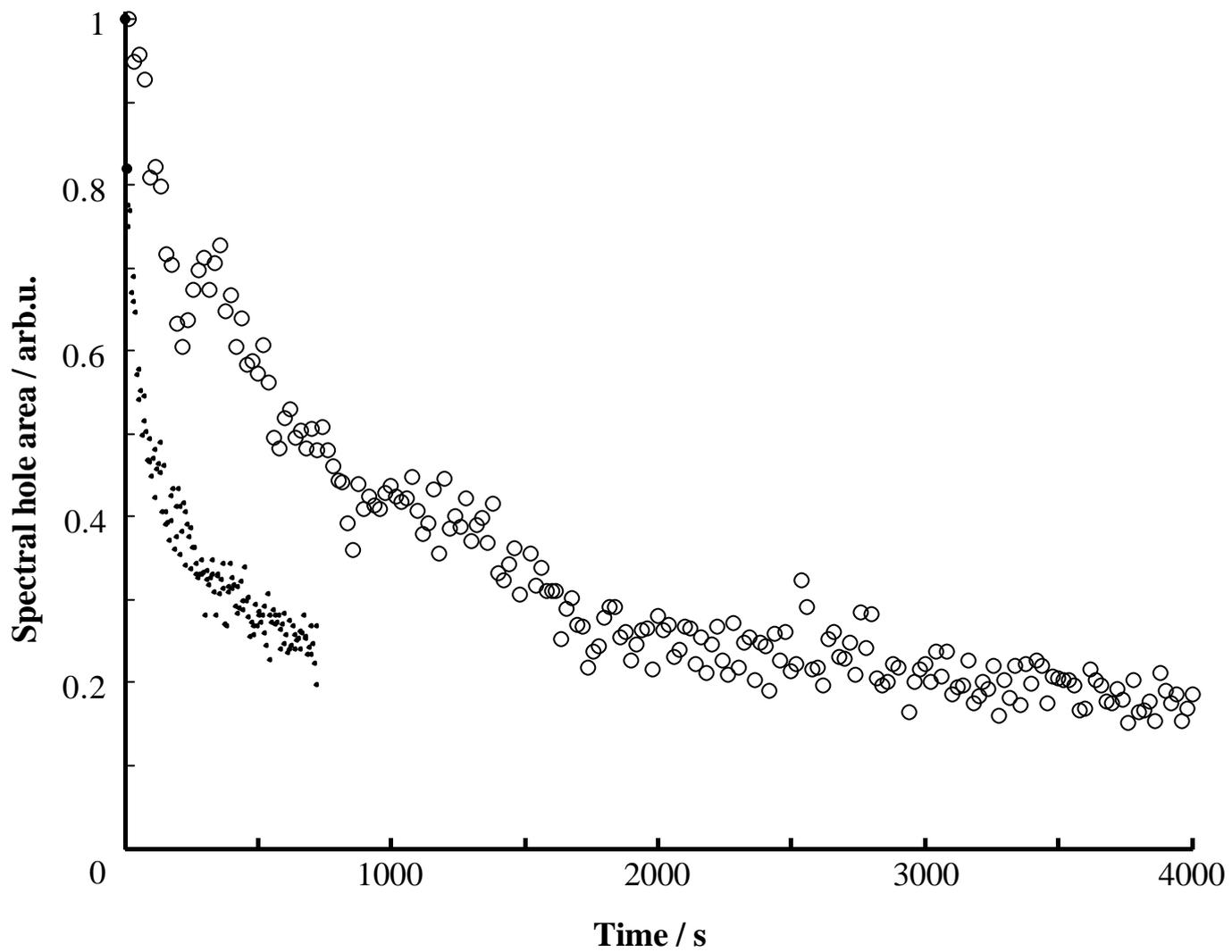

FIG **4**

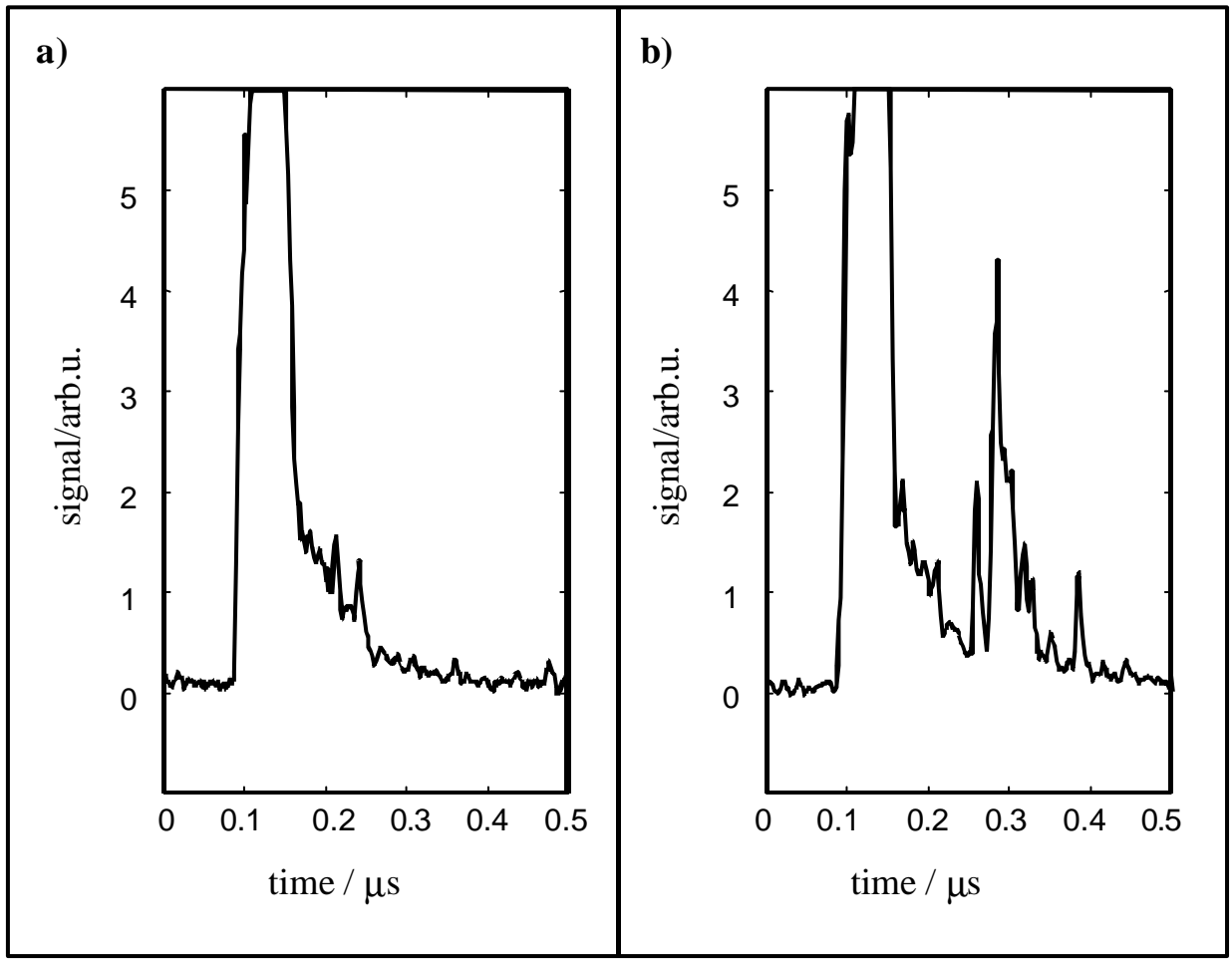

FIG **5**

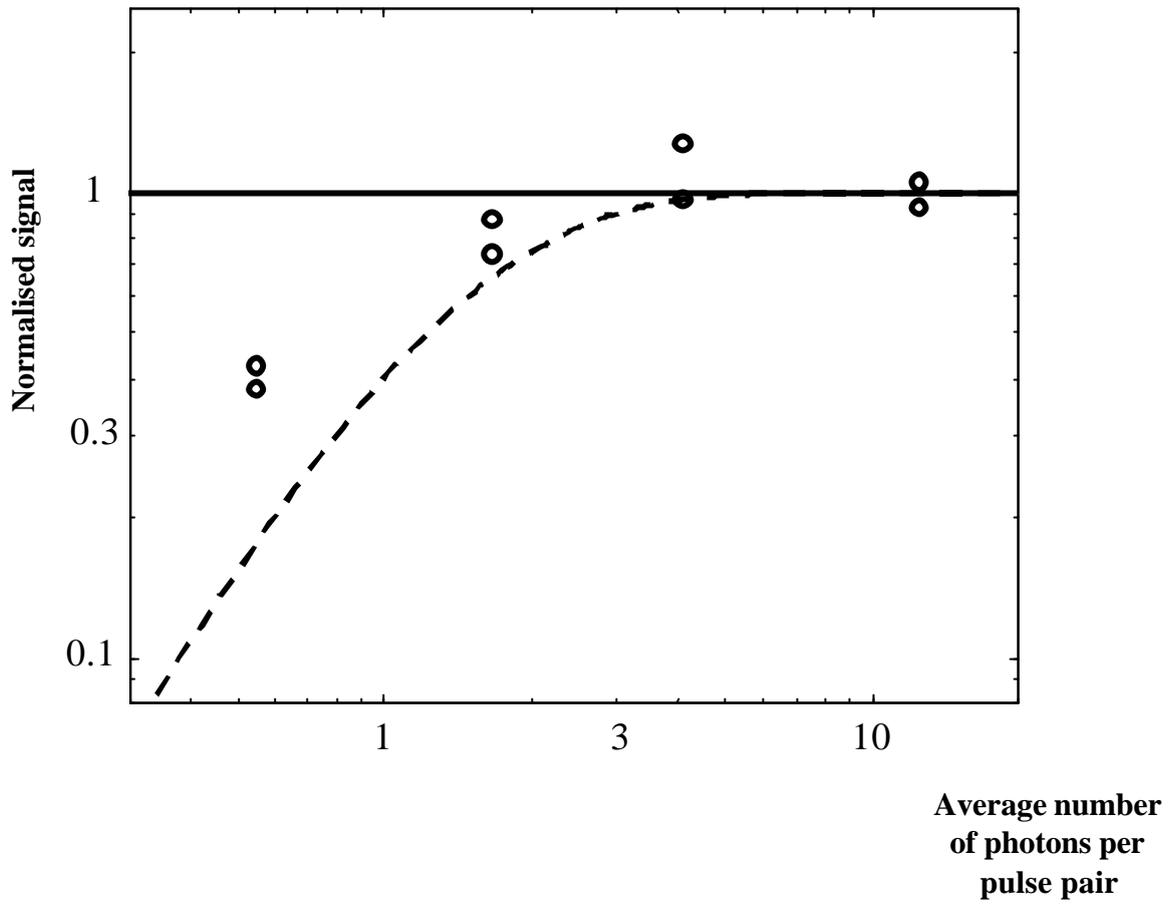

FIG **6**

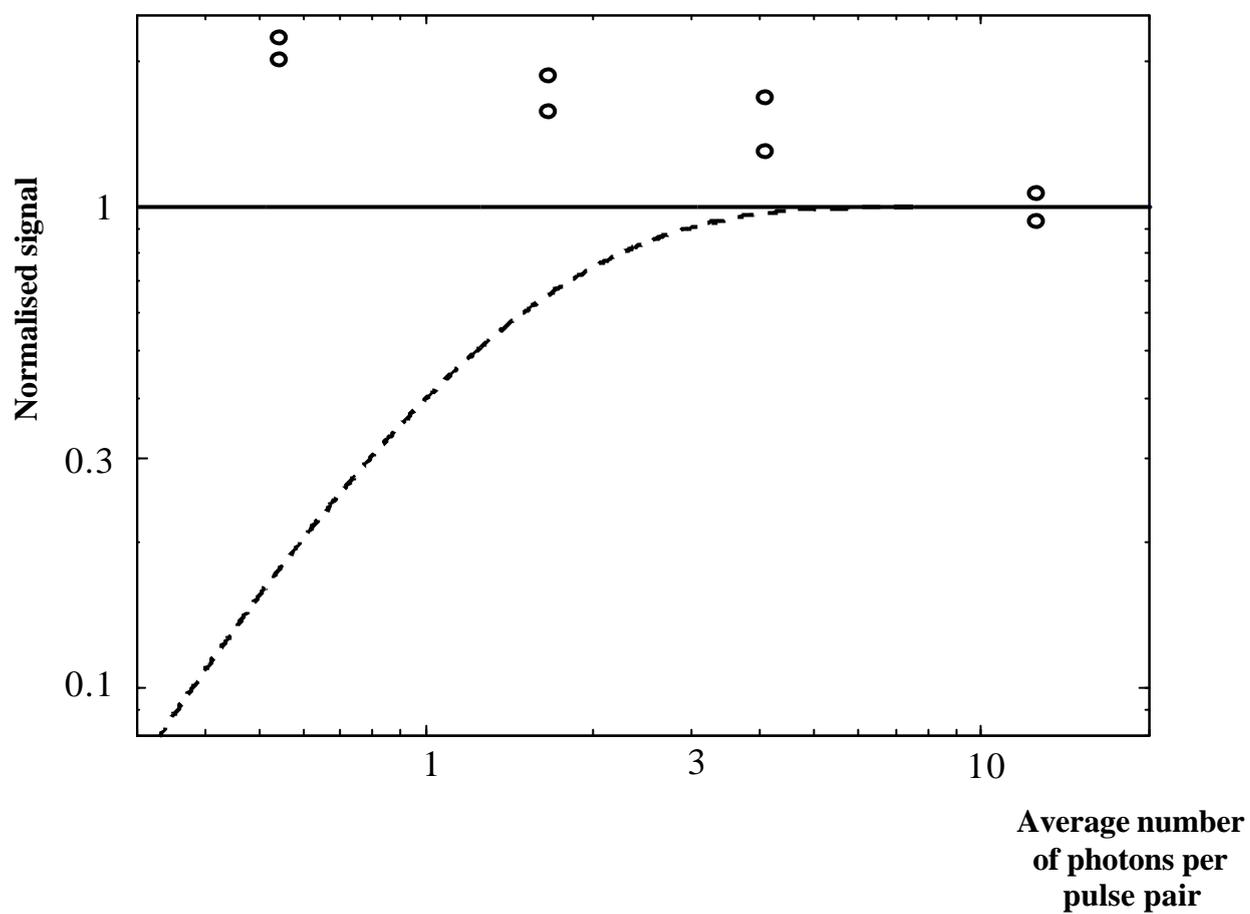

FIG **7**